\documentstyle[12pt]{article}
\newcounter{dummy}{}
\newcommand{\letters}{\setcounter{dummy}{\value{equation}}
\refstepcounter{dummy}
\setcounter{equation}{0}
\renewcommand{\theequation}{\arabic{dummy}\alph{equation}}}
\newcommand{\noletters}{\setcounter{equation}{\value{dummy}}
\renewcommand{\theequation}{\arabic{equation}}}
\newenvironment{mathletters}{\letters}{\noletters}
\date{}
\title{Instantons in curvilinear coordinates}
\author{A.~A.~Abrikosov,~jr. %
\thanks{The work is done with partial support of the RFBR grant
97-02-16131.} \\
{\em ITEP, B.~Cheremushkinskaya str., 25,}   \\
{\em 117218 Moscow, RUSSIA}}
\begin{document}
\maketitle

\begin{abstract}
The multi-instanton solutions by 'tHooft and Jackiw, Nohl \& Rebbi are
generalized to curvilinear coordinates. Expressions can be notably
simplified by the appropriate gauge transformation.  This generates the
compensating addition to the gauge potential of pseudoparticles.
Singularities of the compensating connection are irrelevant for physics
but affect gauge dependent quantities.
\end{abstract}

\section*{The third connection}

The years that passed since the discovery of instantons, \cite{BPST}, did
not bring answer to the question about the role of instantons in QCD,
\cite{Shifman,Schaefer/Shuryak}.  As far as confinement remains a puzzle
all references to instantons at long scales are ambiguous. Indications may
come from studies of instanton effects in phenomenological models.  These
could tell whether confinement may seriously affect pseudoparticles and
{\em v.~v.}

Common confinement models look most natural in non-Cartesian coordinate
frames. The obvious choice for bags are 3+1-cylindrical, {\em i.~e.\/}
3-sphe\-ri\-cal+time, coordinates while strings would prefer
2+2-cy\-lin\-dri\-cal (2+1-cy\-lin\-dri\-cal+time) geometry. Nevertheless
instantons were usually discussed in the Cartesian frame (that was ideal
in vacuum).  The purpose of the present work is to draw attention to the
problem and to develop the adequate technique. We shall generalize to
curvilinear coordinates the multi-instanton solutions by 'tHooft and
Jackiw, Nohl \& Rebbi, \cite{JNR}, and simplify the formulae by the gauge
transformation. Presently I don't know whether the procedure is
good for other topological configurations$^{\dag}$ but I
would expect that it makes sense for the AHDM$^{\ddag}$, \cite{AHDM},
solution%
\footnote[1]{I'm grateful to  L.~Lipatov$^{\dag}$ and
S.~Moch$^{\ddag}$ for the questions.}.%$

We start from the basics of curvilinear coordinates in Sect.~{\ref{cc}}
and introduce the first two connections, namely the Levi-Civita connection
and the spin connection. In Sect.~{\ref{inst}} we describe the
multi-instanton solutions. In Sect.~{\ref{mult-inst}} we shall rewrite
instantons in non-Cartesian coordinates and propose the gauge transform
that makes formulae compact. The price will be the appearance of the {\em
third, so called compensating,\/} gauge connection. The example of the
$O(4)$-sphe\-ri\-cal coordinates is sketched in Sect.~{\ref{examp}}.
Singularities of the gauged solution are discussed in Sect.~{\ref{sing}}.
The last part summarizes the results.

\section{Basics}   \label{basics}

\subsection{Curvilinear coordinates}  \label{cc}

We shall consider flat 4-dimensional euclidean space-time that may be
parametrized either by the set of Cartesian coordinates $x^\mu $ or by
curvilinear ones called $q^\alpha $. The $q$-frame is characterized by the
metric tensor $g_{\alpha \beta }(q)$:
\begin{equation}
ds^2=dx_\mu ^2=g_{\alpha \beta }(q)\,dq^\alpha \,dq^\beta .
\end{equation}

In the $q$-frame the derivatives $\frac \partial {\partial x^\mu }$ should
be replaced by the covariant ones, $D_\alpha $. For example the derivative
of a covariant vector $A_\beta $ is:
\begin{equation}
D_\alpha A_\beta =\partial _\alpha A_\beta
-\Gamma _{\alpha \beta}^\gamma \,A_\gamma .
\end{equation}
The function $\Gamma _{\beta \gamma }^\alpha$ is called the
{\bf Levi-Civita connection.} It can be expressed in terms of the metric
tensor ($g_{\alpha \beta } \, g^{\beta \gamma }=\delta _\alpha ^\gamma $):
\begin{equation}
\Gamma _{\beta \gamma }^\alpha =
\frac 12 g^{\alpha \delta }
\left( \frac{\partial g_{\delta \beta }}{\partial q^\gamma }
+\frac{\partial g_{\delta \gamma }}{\partial q^\beta }
-\frac{\partial g_{\beta \gamma }}{\partial q^\delta }\right) .
\label{Gamma}
\end{equation}

Often it is convenient to use instead of $g_{\alpha \beta }$ the four
vectors $e_\alpha ^a$ called the vierbein:
\begin{equation}
g_{\alpha \beta }(q)=\delta_{ab} \, e_\alpha ^a(q) \,e^b_\beta (q).
\label{g<=>pi}
\end{equation}
Multiplication by $e_\alpha ^a$ converts coordinate (Greek) indices into the
vierbein (Latin) ones,
\begin{equation}
A^a=e_\alpha ^a\,A^\alpha .
\end{equation}

Covariant derivatives of quantities with Latin indices are defined in terms
of the {\bf spin connection }$R_{\alpha \,b}^a(q)$,
\begin{equation}
D_\alpha A^a=\partial _\alpha A^a+R_{\alpha \,b}^a\,A^b.
\end{equation}

The two connections $\Gamma _{\alpha \delta }^\beta $ and $R_{\alpha \,b}^a$
are related to each other as follows:
\begin{equation}
R_{\alpha \,b}^a=e_\beta ^a\,\partial _\alpha e_b^\beta +e_\beta ^a\,\Gamma
_{\alpha \gamma }^\beta \,e_b^\gamma =e_\beta ^a\,(D_\alpha e^\beta )_b.
\label{R_alpha}
\end{equation}

\subsection{Instantons}    \label{inst}

We shall discuss pure euclidean Yang-Mills theory with the $SU(2)$ gauge
group. The vector potential is $\hat{A}_\mu =\frac 12\tau ^a\,A_\mu ^a$
where $\tau ^a$ are the Pauli matrices. The (Cartesian) covariant
derivative is $D_\mu =\partial _\mu -i\,\hat{A}_\mu $, and the action has
the form:
\begin{equation}
S=\int \frac{{\rm tr}\, \hat{F}_{\mu \nu }^2}{2g^2}\,d^4x=
\int \frac{{\rm tr}\,\hat{F}_{\alpha \beta }\,
\hat{F}^{\alpha \beta }}{2g^2}\,\sqrt{g} \, d^4q.
\label{s_gauge}
\end{equation}
where $g=\det \left| \left| g_{\alpha \beta }\right| \right|$. The formula
for the gauge field strength $\hat{F}_{\alpha \beta }$ is universal:
\begin{equation}
\hat{F}_{\alpha \beta }(\hat{A})=\partial _\alpha \hat{A}_\beta -\partial
_\beta \hat{A}_\alpha -i\,\left[ \hat{A}_\alpha ,\,\hat{A}_\beta \right] .
\label{[DmuDnu]}
\end{equation}
The action is invariant under gauge transformations,
\begin{equation}
\hat{A}_\mu \rightarrow \hat{A}_\mu ^\Omega =\Omega ^{\dagger }\,\hat{A}_\mu
(x)\,\Omega +i\,\Omega ^{\dagger }\,\partial _\mu \Omega \,,
\label{A_s=OA_rO+OdO}
\end{equation}
where $\Omega $ is a unitary $2\times 2$ matrix, $\Omega ^{\dagger }=\Omega
^{-1}.$

The field equations have selfdual ($F_{\mu \nu }=\tilde{F}_{\mu \nu }=\frac
12\epsilon _{\mu \nu \lambda \sigma }\,F^{\lambda \sigma }$) solutions known
as instantons. The most general explicit selfdual configuration found by
Jackiw, Nohl and Rebbi, \cite{JNR}, is:
\begin{equation}
\hat{A}_\mu (x)=-\frac{\hat{\eta}_{\mu \nu }^{-}}2\partial _\nu \ln \rho (x),
\label{JNR}
\end{equation}
where $\hat{\eta}_{\mu \nu }$ is the matrix version of the 'tHooft's
$\eta$-sym\-bol, \cite{tHooft}:
\begin{equation}
\hat{\eta}_{\mu \nu }^{\pm }=-\hat{\eta}_{\nu \mu }^{\pm }=\left\{
\begin{array}{cc}
\tau ^{a\,}\epsilon ^{a\mu \nu }; & \mu ,\nu =1,2,3; \\
\pm \tau ^a\,\delta ^{\mu a}; & \nu =4.
\end{array}
\right.   \label{eta^pm}
\end{equation}

The widely used regular and singular instanton gauges as well as the
famous 'tHooft's {\em Ansatz\/} may be cast into the form similar to
(\ref{JNR}).

Our aim is to generalize the solution (\ref{JNR}) to curvilinear
coordinates. We shall not refer to the explicit form of $\rho (x)$ and the
results will be applicable to all the cases.

\section{Multi-instantons in curvilinear coordinates} \label{mult-inst}

\subsection{The problem and the solution}

It is not a big deal to transform the covariant vector $\hat{A}_\mu ,$ (\ref
{JNR}), to $q$-coordinates. However this makes the constant
numerical tensor $\hat{\eta}_{\mu \nu }$ coordinate-dependent
\begin{equation}
\hat{\eta}_{\mu \nu }\rightarrow \hat{\eta}_{\alpha \beta }=\hat{\eta}_{\mu
\nu }\frac{\partial x^\mu }{\partial q^\alpha }\frac{\partial x^\nu }{%
\partial q^\beta }.  \label{eta_x->eta_q}
\end{equation}

We propose to factorize the coordinate dependence by means of the gauge
transformation such that
\begin{equation}
\Omega ^{\dagger }\,\hat{\eta}_{\alpha \beta }\Omega
=e_\alpha ^a\,e_\beta ^b\, \hat{\xi}_{ab}.
\label{Oxi=etaO}
\end{equation}
Here $\hat{\xi}_{ab}$ is a constant numerical matrix tensor,

\begin{equation}
\hat{\xi}_{ab}=\delta _a^\mu \,\delta _b^\nu \,\hat{\eta}_{\mu \nu }.
\label{xi=eta}
\end{equation}
It takes the place of $\hat{\eta}_{\mu \nu }$ in non-Cartesian
coordinates.

It can be shown that the matrix $\Omega $ does exist provided that the
$x$-frame and the $q$-frame have the same orientation and the two sides of
(\ref{Oxi=etaO}) are of same duality.

The gauge-rotated instanton field is the sum of the two pieces:
\begin{equation}
\hat{A}_\alpha ^\Omega (q)=
-\frac 12 e_\alpha ^a\,\hat{\xi}_{ab}\,e^{b\,\beta}\,
\partial _\beta \ln \rho \,(q)
+i\,\Omega ^{\dagger }\,\partial _\alpha \Omega.
\label{A^I+A^comp}
\end{equation}
The first addend is almost traditional and does not depend on the
$\Omega $-matrix whereas the second one carries information about the
$q$-frame. It is entirely of geometrical origin. We call it the {\bf
compensating connection} because it compensates the coordinate dependence
of $\hat{\eta}_{ab}=e_a^\alpha\, e_b^\beta\, \hat{\eta}_{\alpha\beta}$ and
reduces it to the constant $\hat{\xi}_{ab}$.

So long we did not specify what was the duality of the $\hat{\eta}$-symbol.
However the $\Omega $-matrices and compensating connections for
$\hat{\eta}^{+}$ and $\hat{\eta}^{-}$ are different. In general
$\hat{A}_\alpha ^{{\rm comp\,}\pm }$ are respectively the selfdual and
antiselfdual projections of the spin connection onto the gauge group:
\begin{equation}
\hat{A}_\alpha ^{{\rm comp\,}\pm }=i\,\Omega _{\pm }^{\dagger }\,\partial
_\alpha \Omega _{\pm }=-\frac 14\,R_\alpha ^{ab}\,\hat{\xi}_{ab}^{\pm }.
\label{A^comp}
\end{equation}

The last formula does not contain $\Omega $ that has dropped out of the
final result. In order to write down the multi-instanton solution one
needs only the vierbein and the associated spin connection.

\subsection{Triviality of the compensating field.}

The fact of the compensating connection $\hat{A}^{\rm comp}$,
(\ref{A^comp}), being a pure gauge is specific to the flat space.
It turns out that the field strength $\hat{F}_{\alpha \beta }(\hat{A}^{{\rm
comp}})$ is related to the Riemann curvature of the space-time
$R_{\,\alpha \beta }^{\quad \gamma \delta }$:
\begin{equation}
\hat{F}_{\alpha \beta }(\hat{A}^{{\rm comp\,}\pm })
=-\frac 14 R_{\alpha
\beta }^{\quad \gamma \delta }\hat{\xi}_{\gamma \delta }^{\pm }.
\label{F=R/4}
\end{equation}
Thus $\hat{F}_{\alpha \beta }^{}(\hat{A}^{{\rm comp}})=0$ provided that
$R_{\,\alpha \beta }^{\quad \gamma \delta }=0$. Simple changes of variables
$x^\mu \rightarrow q^\alpha $ do not generate curvature and
$\hat{A}^{\rm comp}$ is a pure gauge. However this is not the case in
curved space-times.

\subsection{Duality and topological charge}

As long as we limit ourselves to identical transformations the
vector potential (\ref{A^I+A^comp}) must satisfy the classical field
equations. However the duality equation looks differently in non-Cartesian
frame. If written with coordinate indices it is:
\begin{mathletters}
\begin{equation}
\hat{F}_{\alpha \beta }=
\frac{\sqrt{g}}2\,\epsilon _{\alpha \beta \gamma \delta }\,
\hat{F}^{\gamma \delta }.
\label{dual-greek}
\end{equation}
Still it retains the familiar form in the vierbein notation:
\begin{equation}
\hat{F}_{ab} =  \frac 12\, \epsilon _{abcd}\,\hat{F}^{cd}.
\label{dual-lat}
\end{equation}
\end{mathletters}

The topological charge is given by the integral
\begin{mathletters}
\label{q}
\begin{equation}
q=\frac 1{32\pi ^2} \int \epsilon _{\alpha \beta \gamma \delta }\,
{\rm tr\,}\hat{F}^{\alpha \beta }\,\hat{F}^{\gamma \delta }\,d^4q,
\label{q-greek}
\end{equation}
which in the vierbein notation becomes
\begin{equation}
q=\frac 1{32\pi ^2}\int \epsilon _{abcd}\,
{\rm tr\,}\hat{F}^{ab}\,\hat{F}^{cd}\,\sqrt{g}\,d^4q.
\label{q-lat}
\end{equation}
\end{mathletters}

The general expression for $\hat{F}_{\alpha \beta }$ in non-Car\-te\-sian
frame is rather clumsy but it simplifies for one instanton. The vector
potential in regular gauge is, ($r^2=x_\mu ^2$):
\begin{equation}
\hat{A}_\mu ^I=
\frac{\hat{\eta}_{\mu \nu }^+}2
\partial _\nu \ln \left(r^2+\rho ^2\right) .
\label{A_reg}
\end{equation}
The conjugated coordinate and $\Omega _{+}$ gauge transformations convert
it into
\begin{equation}
\hat{A}_\alpha ^I=\frac 12e_\alpha ^a\,\hat{\xi}_{ab}\,e^{b\,\beta
}\,\partial _\beta \ln ( r^2+\rho ^2) + \hat{A}^{{\rm comp}\,+},
\end{equation}
and the field strength becomes plainly selfdual:
\begin{equation}
\hat{F}_{ab}(\hat{A}_{}^{I\,})=
-\frac{2\,\hat{\xi}_{ab}^{+}}{\left( r^2+\rho ^2\right) ^2},
\label{F_reg}
\end{equation}
This generalizes regular gauge to any non-Car\-te\-sian coordinates.

\section{Example}  \label{examp}

We shall consider an instanton placed at the origin of the $O(4)$-spherical
coordinates. Those are the radius and three angles:
$q^\alpha =(\chi ,\,\phi ,\,\theta ,\,r)$. The polar axis is aligned with
$x^1$ and
\begin{mathletters}
\begin{eqnarray}
x^1 &=&r\cos \chi ; \\
x^2 &=&r\sin \chi \sin \theta \cos \phi ; \\
x^3 &=&r\sin \chi \sin \theta \sin \phi ; \\
x^4 &=&r\sin \chi \cos \theta .
\end{eqnarray}
\end{mathletters}
The vierbein and the metric tensor are diagonal:
\begin{equation}
e_\alpha ^a=
{\rm diag}\,(r,\,r\sin \chi \sin \theta ,\,r\sin \chi ,\,1).
\end{equation}

Now one may start from the vector potential (\ref{A_reg}) and consecutively
carry out the entire procedure. But to the calculation of the instanton part
this includes calculating $\Gamma _{\beta \gamma }^\alpha $, finding
$R_\alpha ^{ab}$ and, finally, computing $\hat{A}^{{\rm comp}\,+}$. The
$\hat\xi_{ab}^{+}$-symbol coincides with $\hat{\eta}_{ab}^{+}$,
(\ref{eta^pm}, \ref{xi=eta}). The result is:
\begin{mathletters}
\label{instO(4)}
\begin{eqnarray}
\hat{A}_\chi ^I &=&
\frac{\tau _x}2\left( \frac{r^2-\rho ^2}{r^2+\rho ^2} \right) ;
\label{instO(4)a} \\
\hat{A}_\phi ^I &=&
-\frac{\tau _x}2\cos \theta +
\frac{\tau _y}2\sin \chi \sin \theta
\left( \frac{r^2-\rho ^2}{r^2+\rho ^2}\right)   \nonumber \\
&&+\frac{\tau _z}2\cos \chi \sin \theta ;
\label{instO(4)b} \\
\hat{A}_\theta ^I &=&
-\frac{\tau _y}2 \cos \chi +
\frac{\tau _z}2\sin \chi
\left( \frac{r^2-\rho ^2}{r^2+\rho ^2}\right) ;
\label{instO(4)c} \\
\hat{A}_r^I &=&0.  \label{instO(4)d}
\end{eqnarray}
\end{mathletters}
The corresponding field strength is given by (\ref{F_reg}).

\newpage
\section{Singularities}    \label{sing}

Note that the vector field (\ref{instO(4)}) is singular since neither
$\hat{A}_\theta ^I$ nor $\hat{A}_\phi ^I$ goes to zero near polar axes
$\chi =0$ and $\theta =0$. These singularities are produced by the
gauge transformation and must not affect observables. However they may
tell on gauge variant quantities. We shall demonstrate that for the
Chern-Simons number.

The topological charge, (\ref{q}), can be represented by the surface
integral, $q=\oint K^\alpha \,dS_\alpha $,
\cite{Shifman,Schaefer/Shuryak}. Here,
\begin{equation}
K^\alpha =
\frac{\epsilon ^{\alpha \beta \gamma \delta }}{16\pi ^2}
\,{\rm tr\,}\left( \hat{A}_\beta \,\hat{F}_{\gamma \delta }
+\frac{2i}3\hat{A}_\beta \hat{A}_\gamma \hat{A}_\delta \right) .
\end{equation}
Even though $q$ is invariant $K^\alpha $ depends on gauge. Consider
Cartesian instanton in the $\hat{A}_4=0$ gauge. The two contributions to the
topological charge come from the $x_4=\pm \infty $ hyperplanes,
$q= N_{\rm CS}(\infty)- N_{\rm CS}(-\infty)$, and the quantity
\begin{equation}
N_{\rm CS}(t)=\int_{x_4=t}K^4\,dS_4
\end{equation}
is  called the Chern-Simon number. Instanton is a transition between two
3-di\-men\-sional vacua with $\Delta N_{{\rm CS}}=1$.

Analysis of (\ref{instO(4)}) reveals a striking resemblance with this case.
By coincidence here again $\hat{A}_4=0,$ (\ref{instO(4)d}). This
gives an idea to interpret $r$ as a time coordinate attributing the
Chern-Simons number $N_{{\rm CS}}(r)$ to the sphere of radius $r$. A naive
expectation would be that $\Delta N_{{\rm CS}}=\left. N_{{\rm CS}}(r)\right|
_0^\infty $ gives the topological charge. However this is not true and
$\Delta N_{{\rm CS}}=\frac 12$. The second half of $q$ is contributed by the
singularities at $\theta =0,\,\pi $. The $\Omega $-transform has affected
the distribution of $N_{{\rm CS}}$.

We conclude that in our approach gauge variant quantities depend on
coordinate frame and may be localized at the singularities of the
$\Omega$-transform. This may be one more way to simplify calculations with
the help of curvilinear coordinates.

\section*{Summary}

We have shown that explicit {\em (multi-)\/}instanton solutions can be
generalized to curvilinear coordinates. The gauge transformation converts
the coordinate-dependent $\hat{\eta}_{ab}$-symbol into the constant
$\hat{\xi}_{ab}$. The gauge potential is a sum of the instanton part and
the compensating gauge connection, (\ref{A^I+A^comp}).

The compensating gauge connection can be computed in the three steps:
\begin{enumerate}
\item  One starts from the calculation of the Levi-Civita connection
$\Gamma _{\beta \gamma }^\alpha $, (\ref{Gamma}).
\item  Covariant differentiation of the vierbein, (\ref{R_alpha}), leads
to the spin connection $R_\alpha ^{ab}$.
\item  Convolution of the spin connection with the appropriate
$\hat{\xi}_{ab}$ gives the compensating gauge potential, (\ref{A^comp}).
\end{enumerate}

The advantage of our solution is that it is constructed directly of
geometrical quantities, {\em i.~e.\/} the vierbein and the spin
connection. Another attractive feature is the relation between gauge
variant quantities and the coordinate frame. More details may be found in
\cite{9906008}.

\section*{Acknowledgments}

I would like to acknowledge financial support of the Ministry of Science of
Russian Federation as well as that of the Organizing Committee which made
possible my participation in the conference. It is a pleasure to thank the
Chairman, professor S.~Narison and the staff for personal care of
participants.

%----------- References ----------------------------------

\end{document}